\documentstyle[11pt,paspconf,epsf]{article}

\begin{document}

\title{IFIRS: An Imaging Fourier Transform Spectrometer for 
the Next Generation Space Telescope}

\author{James. R. Graham}
\affil{Department of Astronomy, University of California, Berkeley, CA 94720}

\begin{abstract}

Due to its simultaneous deep imaging and integral field spectroscopic
capability, an Imaging Fourier Transform Spectrograph (IFTS) is
ideally suited to the Next Generation Space Telescope (NGST) mission,
and offers opportunities for tremendous scientific return in many
fields of astrophysical inquiry.  We describe the operation and
quantify the advantages of an IFTS for space applications. The
conceptual design of the Integral Field Infrared Spectrograph (IFIRS)
is a wide field ($5'.3 \times 5'.3$) four-port imaging Michelson
interferometer.
\end{abstract}

\section{Introduction}

\begin{figure}[t]
\plotfiddle{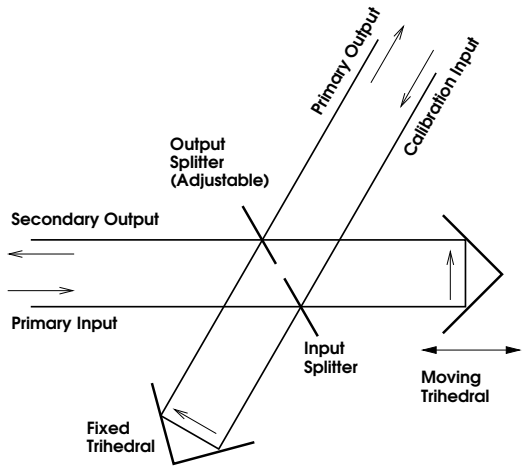}{2.0in}{0}{35}{35}{-120}{0}
\caption{\small An IFTS is a Michelson interferometer where the
telescope focal plane is imaged onto a detector array. An
interferogram is recorded for every pixel, and hence a spectrum can be
obtained for every object. IFIRS has a four-port design
that wastes none of the light. IFIRS uses
cube-corners, which displace the input
and output beams.}
\label{imaging.michelson}
\end{figure}

IFIRS is a Michelson interferometer configured as an imaging Fourier
transform spectrometer (IFTS) (Graham et al. 1998; Graham et
al. 2000).  Fig.~\ref{imaging.michelson} shows a moving mirror which
introduces an optical path difference (OPD) between the two beams
created by the beam-splitter; the resulting interferogram from the
combined beams is recorded for every pixel in the field of view, and
hence a spectrum is obtained for every object.  Since the bandpass of
the instrument is defined only by the detectors and the efficiency of
the beam-splitter, broad wavelength coverage is an intrinsic feature
of the IFTS.  An IFTS has continuously variable spectral resolution up
to a maximum defined by the maximum OPD, which in the IFIRS design is
1~cm (i.e., $R_{max}=10,000$ at 1~$\mu$m). IFIRS is a four-port
interferometer, a choice which yields several advantages.  (1)
Virtually all the light collected by the telescope is directed toward
the focal plane for detection.  (2) The final interferogram,
constructed from the difference of the two output ports, is immune to
common mode noise.

When the interferograms from the two output ports are summed, the
total flux image is recovered, producing the simultaneous deep
panchromatic image that is a unique capability of an IFTS.  In
comparison with a simple camera, in which the panchromatic image is
formed by summing individual filter images, the IFTS panchromatic
image has a speed advantage factor equal to the number of filters
used.  In its hybrid, or dispersed FTS mode, designed for use with
slit masks at high spectral resolution ($R\simeq600-10,000$), IFIRS
delivers a sensitivity for each object equivalent to that of a
conventional dispersive spectrograph, with the spatial multiplex
advantage of a multi-object spectrometer. The capabilities
of IFIRS are summarized in Table \ref{capabilities}.

\begin{table}
{\small
\begin{center}
\caption{\bf Observational Capabilities of IFIRS}
\medskip
\begin{tabular}{lll}
\hline
		& Near-IR Channel 	& Mid-IR Channel \\
\hline
Bandpass/Detector	& 0.6-5.6 $\mu$m/InSb & 3-28 $\mu$m/Si:As \\
Maximum Spectral Resolution,$R$ 	& 15,000	& 3000 \\ 
FOV/Array Format	& $5.'3$/8k$\times$8k & $2.'6$/2k$\times$2k \\
Pixel size/Nyquist sampled $\lambda$ 	& $0.''0386$/3 $\mu$m & $0.''0772$/6 $\mu$m  \\
Wavefront error/Strehl		& 150 nm rms/0.8& 150 nm rms/0.99 \\
Throughput (excluding detector) 	& 86\% & 87\% \\
FTS Sensitivity, $R=1/5/100$ & 0.2/1/35 nJy & 13/65/1300 nJy \\
Dispersed FTS Sensitivity, $R=600$ & 66 nJy & 1080 nJy \\
\hline
\end{tabular}
\label{capabilities}\\
Strehl ratios and sensitivities are quoted at
2 \& 10 $\mu$m. $SNR$ = 10, $t=10^5$~s.
\end{center}
}
\end{table}

\section{IFIRS Performance}

\begin{figure}[t]
\plotfiddle{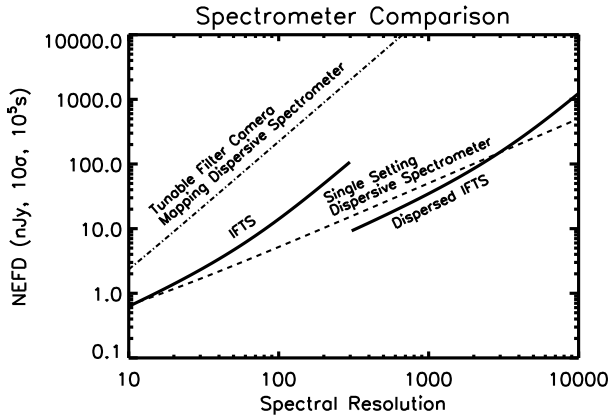}{2.0in}{0}{35}{35}{-140}{0}
\caption{\small The NEFD per pixel for a variety of NGST imaging
spectrometers (Bennett 2000).  All observations extend over
the $K$-band. The IFTS obtains
spectra for every pixel.  For $R > 300$ the IFTS curve is for the
dispersed IFTS mode.
The mapping dispersive
spectrometer is scanned over a number field of view settings, assumed
to be equal to the number of spectral channels acquired by the tunable
filter camera, and thus they have equal NEFD.  The single setting
dispersive spectrometer acquires spectra for a single pointing. The
same NEFD level is achieved by a tunable filter camera if all the
integration time is devoted to observing a single wavelength.  
}
\label{nefd-plot}
\end{figure}

Since NGST must obtain images and spectra of large samples of objects
in order to achieve its science goals, IFIRS's ability to deliver
diffraction-limited full-field imaging spectroscopy is a critical
advantage. Fig.~\ref{nefd-plot} compares the noise equivalent flux
density (NEFD) per pixel as a function of spectral resolution for NGST
backgrounds at 2~$\mu$m for a tunable filter camera, a dispersive
spectrometer, and an IFTS (Bennett 2000).  The IFTS substantially
outperforms both the tunable filter and the dispersive spectrometer
used in mapping mode.  The dispersive spectrometer has the best
performance for spectroscopy, although only for the small number of
objects that lie on the slit.

A hybrid approach (e.g., Beer 1992), which takes advantage of the best
features of all of the 3-d imaging approaches, is the combination of
an objective prism with an IFTS.  Low dispersion in one dimension of
the image plane reduces the spectral bandpass acceptance that is
involved in the noise term for the IFTS.  With a slit at an image
plane, the panchromatic output of the IFTS would yield the same
results as an ordinary prism spectrometer, while the Fourier
transformed interferograms would enable much higher spectral
resolution at much reduced NEFD.  The $R> 300$ segment of the FTS
curve in Fig.~\ref{nefd-plot} is for this hybrid configuration, and
corresponds to the addition of an $R\simeq 600$ dispersing prism and a
focal plane mask.

The curves in Fig.~\ref{nefd-plot} depend on detector performance.
Detectors with higher read noise and dark current would decrease the
sensitivity of the tunable filter, the dispersive spectrometer, and
the dispersed IFTS, but produce little change in the IFTS.

\begin{figure}
\plottwo {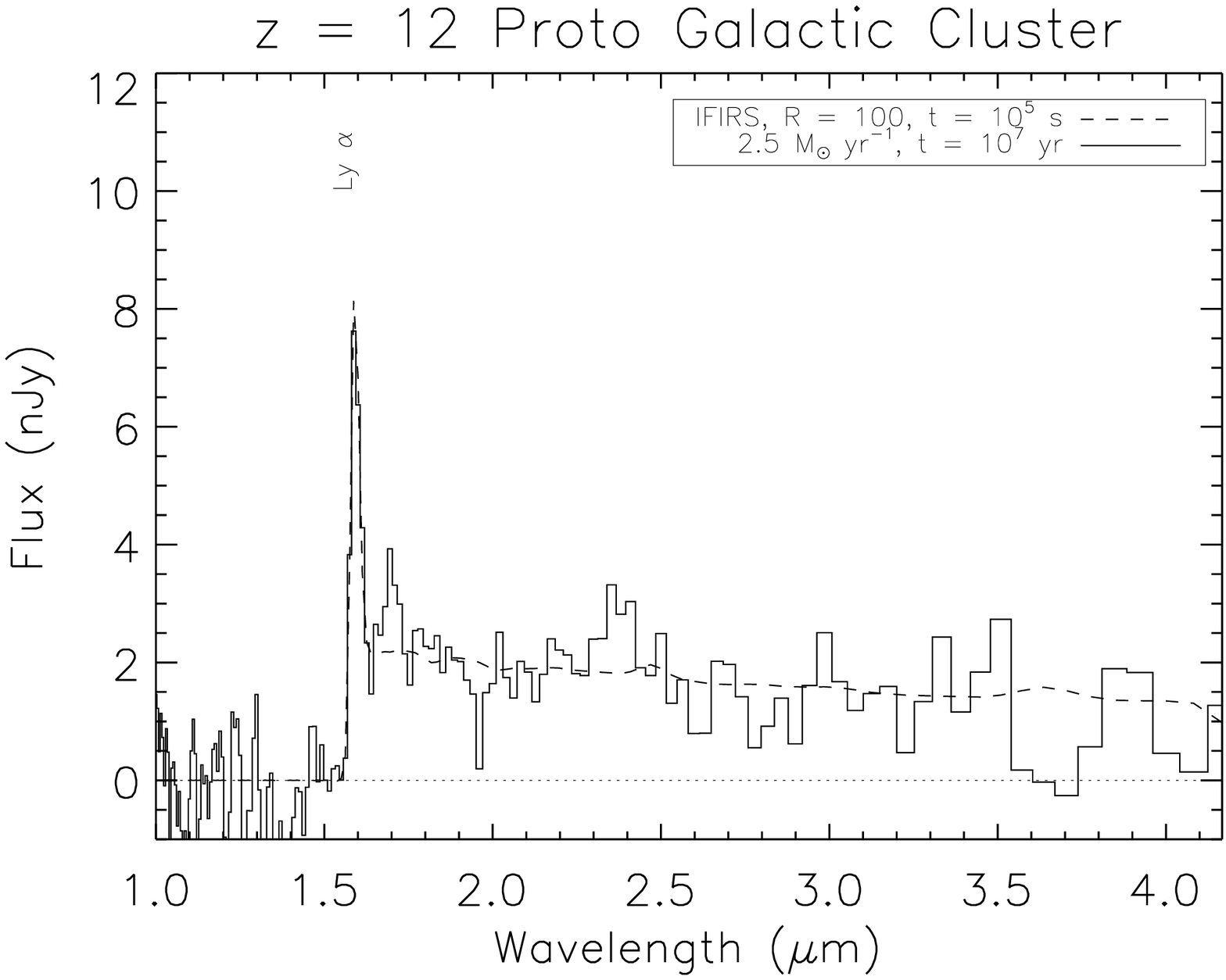}{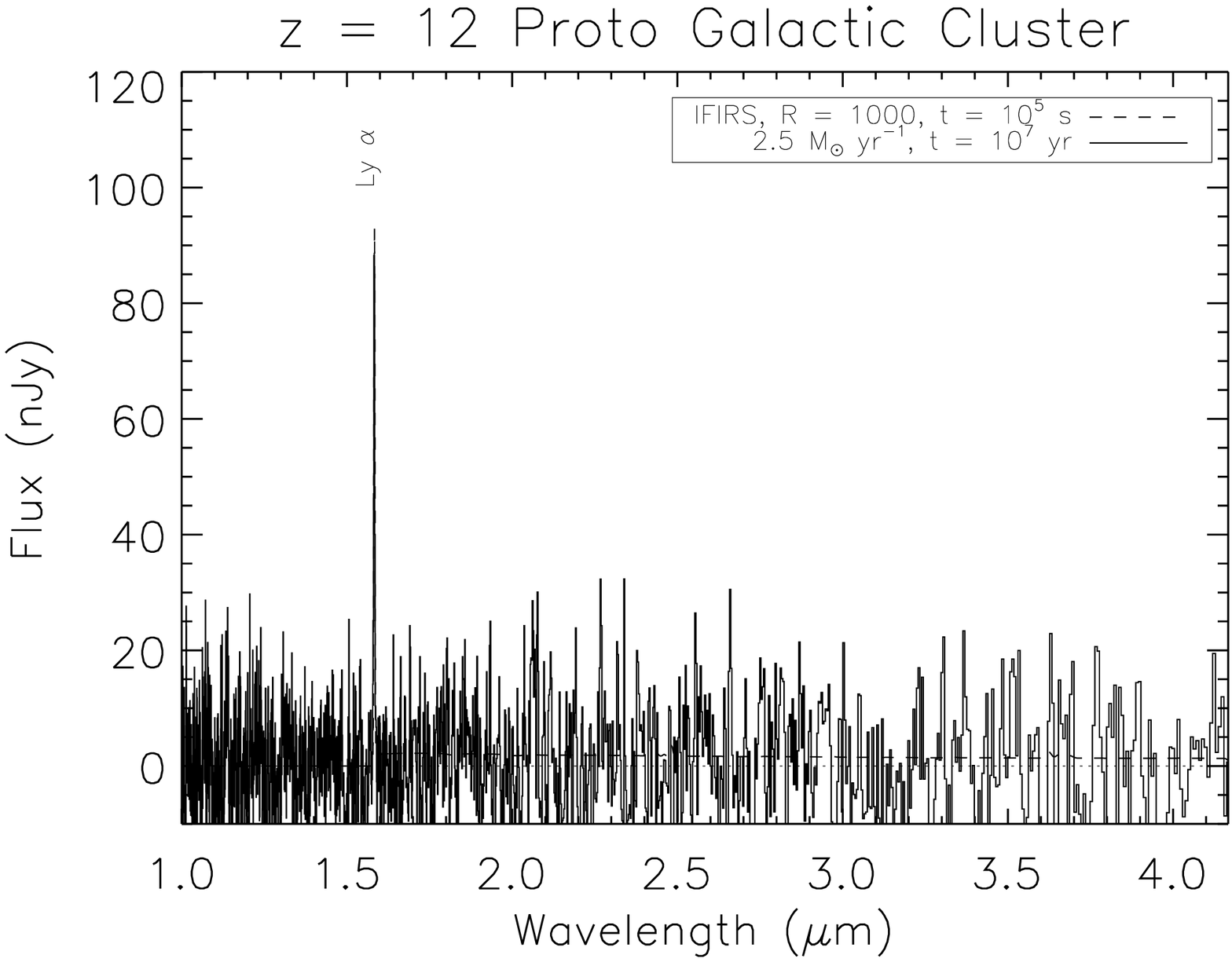}
\caption{\small The sensitivity of an pure IFTS is independent of
spectral resolution. Simulated observations of a $z=12$ star
forming cluster are shown at $R=100$ and $R=1000$. The exposure time
is fixed. Low or high resolution IFTS scans can be used to search for
line emission, or broad spectral features such as the Lyman break.
Kinematics, chemical composition, or peculiar velocities can be
measured in the same exposure time at high resolution.}
\label{variable_resolution}
\end{figure}

One important aspect of the IFTS is that the sensitivity to an
unresolved line is independent of $R$, so long as read noise 
can be
neglected.  Fig.~\ref{variable_resolution} illustrates this by showing
the same star forming galaxy at $R = 100$ and 1000. In each case 
Ly$\alpha$ is detected with the same SNR. In a FTS the noise is
determined by the total photon shot noise, i.e., it depends only on
the total integration time. Therefore, so long as the line is
unresolved, the SNR remains constant. The same is true for a
dispersive spectrometer which is dark current limited, which is
typically the case for NGST. In the dark current limited domain the
dispersive spectrometer looses much of its advantage for high spectral
resolution observations.

\subsection{Spectrometer Comparisons}

Several programs in the NGST design reference mission (DRM) call for
low resolution spectra of large numbers of faint ($K > 29$ AB)
objects. Since dispersive spectrometers on NGST are dark current
limited, they are unsuitable for large sample surveys.  Survey science
is crucial to NGST; for example, one component of the galaxy evolution
DRM consists of validating photometric redshifts with $R \simeq 100$
spectroscopy.  The accuracy and precision of photometric redshifts
should be investigated over a range of galaxy properties.  Since
statistical errors scale as the square root of the number of galaxies
in each bin, of order 100 galaxies are needed per bin.  Without
such large numbers multiplicative factors decimate small samples to
statistical meaninglessness.  A modest sample to test photometric
redshifts consists of 10-redshift $\times$ 5-luminosity $\times$
2-color $\times$ 3-environment $\times$ 3-morphology bins $\times$ 100
galaxies per bin = 90,000 galaxies. The IFTS can perform a complete
magnitude limited sample to $K = 29$~AB in three days. An MOS would
take 45 days to complete this survey.

To discover and characterize the population of sources which reionize
the universe at $z \ga 10$ requires winnowing out sources which
comprise $<$ 0.1 \% of the population at $K \simeq 29-30$ AB (Haiman
\& Loeb 1998).  Clearly thousands of spectra are needed for this task
to: 1) find the high-z population; 2) distinguish AGN from objects
powered by star formation. Once these distant star-clusters, galaxies,
or quasars have been identified, their spectra may be coadded to
construct a very high signal-to-noise ratio composite spectrum that is
a sensitive probe of the re-ionization epoch.  Rather than hoping to
find a single bright target in the dark ages, composite coadded
spectra may be the most reliable way to search for the reionization
epoch: the luminosities and masses of the first collapsed objects are
unknown, and there are likely to be many more small (faint) ones than
large (bright) ones.

These science programs suggest that it is appropriate to define the
figure of merit used to compare a MOS and an IFTS as the ratio of the
number of galaxy spectra obtained.  We assume that a magnitude limited
catalog is obtained in a fixed observing time, and the same wavelength
range and number of spectral channels for each instrument.

For a single object a dispersive spectrometer is faster than the IFTS
by a factor which is proportional to $R$, under the assumption of
photon shot noise limited performance.  Thus, if the integral galaxy
number counts scale as $f_0^{-\alpha}$, then the speed of the IFTS
compared to a MOS is proportional $f_0^{-\alpha}/(N_{slit} R)$ where
$f_0$ corresponds to the magnitude limit of the catalog and $N_{slit}$
is the number of MOS slits (assuming one object per slit).  At faint
enough fluxes the IFTS always wins because the predicted value of
$\alpha \simeq 1$ (Haiman \& Loeb 1998).  The conclusion is
strengthened because dispersive spectrometers on NGST tend to be dark
current limited. The MOS enters the domain of diminishing returns for
targets where dark current dominates the noise

\begin{figure}[t]
\plottwo{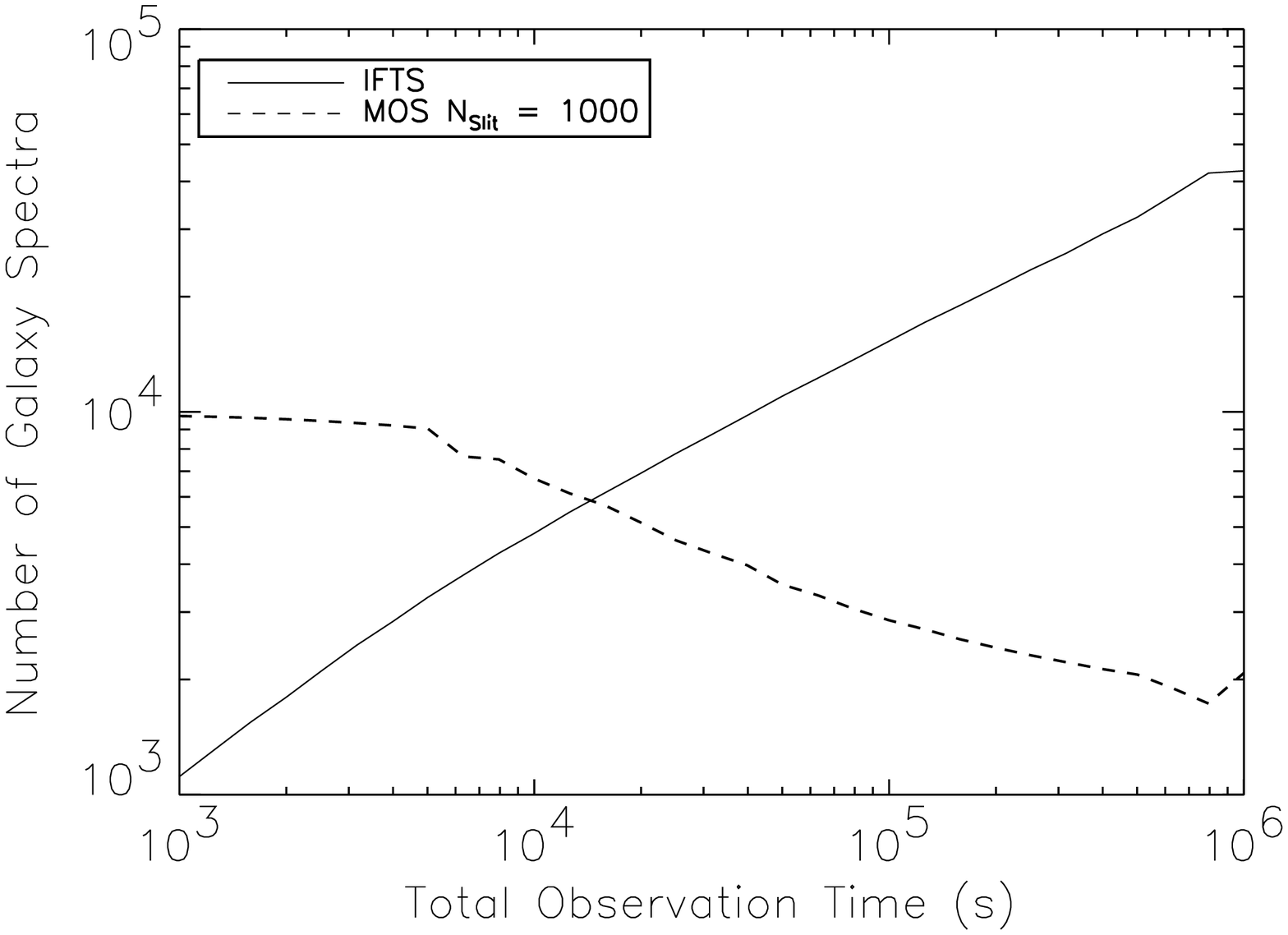}{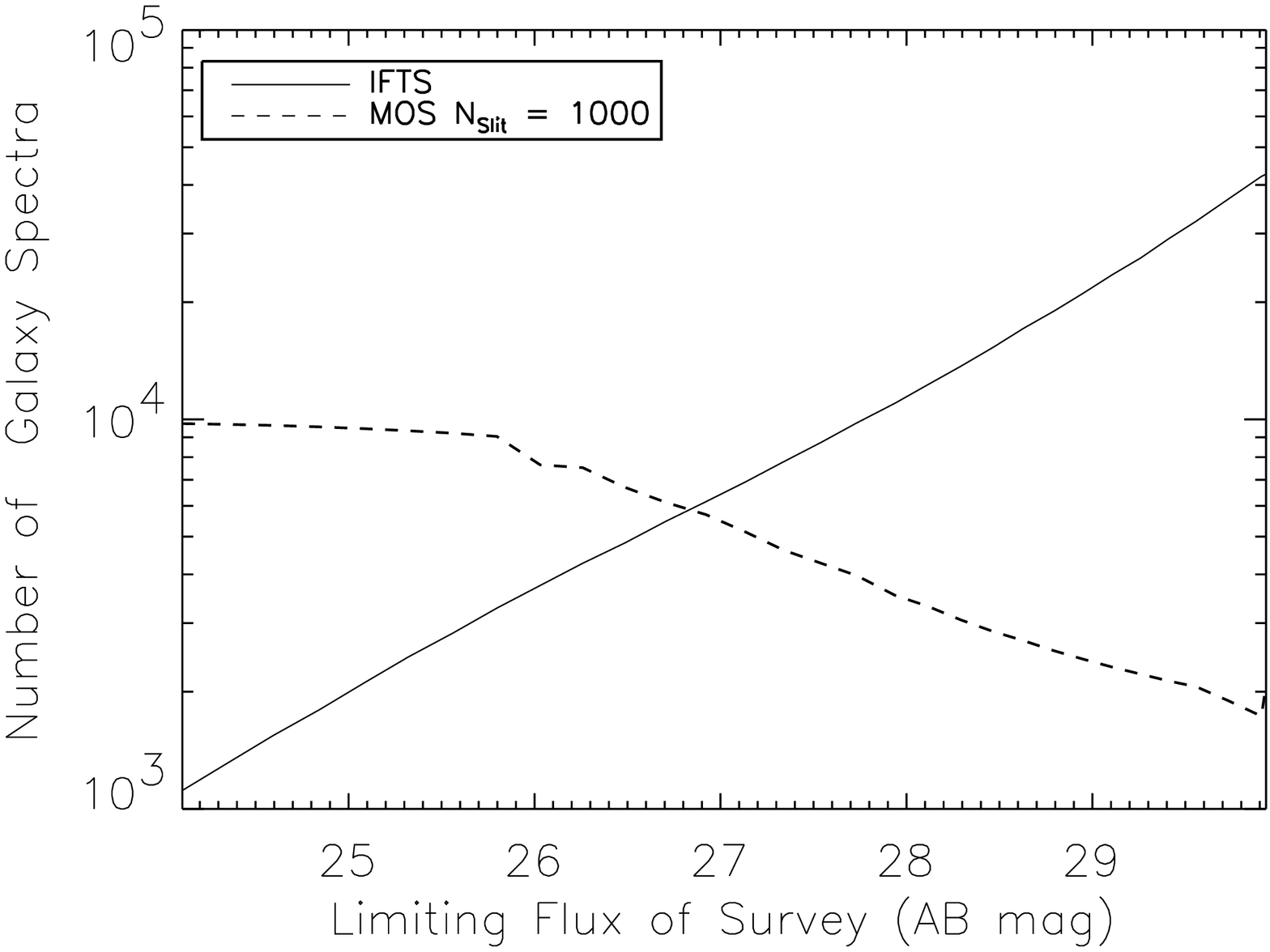}
\caption{\small Comparison of number of galaxy spectra ($SNR > 10$)
obtained as a function of the limiting magnitude and total observation
time for a multislit dispersive spectrometer and a IFTS. The
parameters are: IFTS FOV = $5' \times 5'$; $N_{slit}$ = 1000; $R =
100$; $\Delta \lambda = 1.3 - 2.6$~ $\mu$m (FSR of a 2 $\mu$m blazed
grating, $m=1$); dark current: 0.02 $e^- s^{-1}$; read noise: 4.0
$e^-$ rms; maximum time before readout to veto cosmic rays: 800~s.  }
\label{ftsmoscomparison}
\end{figure}

Fig.~\ref{ftsmoscomparison} shows the results of a more realistic
comparison which plots the total number of galaxy spectra obtained in
a survey.  The IFTS observation is done by using the integration time
to observe one field. For bright objects, the MOS has the speed to
observe multiple fields. Thus the number of MOS spectra equals
$N_{slit}$ times the number of pointings.  The surface density of
galaxies is derived from the Haiman \& Loeb (1998) and Im \& Stockman
simulations and normalized to deep NICMOS counts
(Fig.~\ref{faint-galaxy-counts}).  The two plots which make up 
Fig.~\ref{ftsmoscomparison} are projections of a single 2-d surface which
describes the results of the observation. Either the total observation
time or the limiting magnitude of the survey can be considered as
the independent variable.

For $K < 26$ AB the MOS is photon shot noise limited, and is a fixed
factor faster than the IFTS. The MOS can take data for $\simeq 10$
pointings and obtain spectra for $\simeq$ 10,000 galaxies. Because
such faint galaxies are rare, the IFTS can only obtain spectra for a
few thousand galaxies within its FOV.  For $K > 26$~AB the MOS is dark
current limited, and the speed advantage relative to the IFTS
declines. By $K = 26.9$ AB the MOS and the IFTS acquire equal numbers
of galaxies. As we enter the domain of scientific interest to NGST, $K
> 28$ AB, the IFTS obtains an order of magnitude more galaxy spectra
than the MOS.  Fig. \ref{ftsmoscomparison} shows that the only
practical way to complete the two examples of low resolution
spectroscopy from the DRM is with an IFTS.

\section{IFIRS Science}
\label{Origin}

This brief note only permits description of a few of the NGST science
programs that benefit from an FTS architecture.

\subsubsection{The Origin and Evolution of Galaxies:}
A cornerstone of NGST's science is to understand the formation and
evolution of galaxies.  NGST is uniquely positioned to investigate the
formation of the first galaxy fragments, their growth by merging into
galaxies, and the subsequent chemical and dynamical evolution into the
present day Hubble sequence.  These science goals are addressed by
five conventional imaging and spectroscopic surveys in the DRM.  The
surveys target relatively large samples of galaxies: imaging is
required to determine morphology and select spectroscopic samples, low
resolution $R\simeq100$ spectroscopy of large samples targets global
properties (e.g., redshifts, star formation rate, metallicity, mean
stellar age), and higher resolution $R\simeq1000$ spectroscopy of
smaller samples to determine gas phase abundance, stellar content,
dynamical masses, and dust content.

\begin{figure}[t]
\plotfiddle{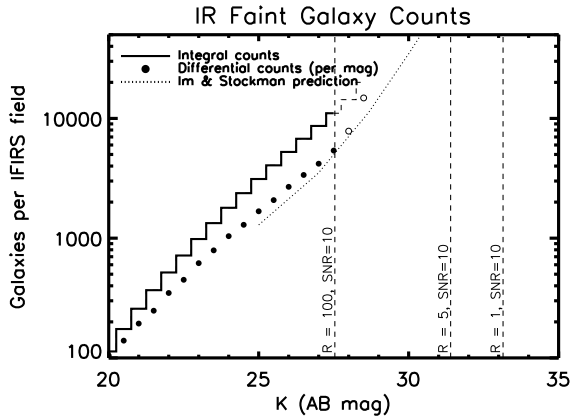}{2.0in}{0}{35}{35}{-140}{0}
\caption{\small Merged galaxy counts from Yan et al. (1998) for $H <
25.9$ AB, Bershady et al. (1998) for $K < 24.9$ AB, Thompson et al
(1999) for $H < 28.8$ AB.  Incompleteness has been corrected according
to Thompson et al.  For $K > 27.5$ AB a dashed line or open symbols
indicate where uncertainties are significant ($ \ga \times 2$) due to
small field observed by NICMOS, photometric errors, and statistical
errors. The IFIRS $SNR = 10$ thresholds for deep exposures ($10^5$~s)
at $R$ = 1, 5, \& 100 at $K$ are indicated by the vertical lines.}
\label{faint-galaxy-counts}
\end{figure}

The steeply rising galaxy number counts
(Fig.~\ref{faint-galaxy-counts}) suggest that the NGST focal plane may
be crowded with $\sim10^5$ faint galaxies per $5.'3 \times 5.'3$ FOV,
i.e., one object per square arcsecond. The efficient study of objects
at high surface densities necessitates a highly multiplexed approach,
such as that afforded by IFIRS.  IFIRS's ability to simultaneously
carry out the imaging and low spectral resolution studies increases
efficiency by combining different science
programs in a single observation and enhances the discovery potential.

To demonstrate the ability of IFIRS to detect and study faint objects
we have simulated data as follows.  An astronomical scene is
represented as a noise-free distribution of objects and associated
spectra. This input data cube is convolved with the telescope point
spread function (PSF).  The spectra are then multiplied by the
wavelength dependent throughput. From this spectral data cube the
interferogram cube is calculated by a Fourier transform, and noise is
added at each OPD step. The noise sources treated are photon shot
noise (from the target, the zodiacal light, and thermal
emission from the telescope), detector noise due to dark current,
and read-noise.  The noisy interferogram cube is then
Fourier transformed back into a spectral data cube.

\begin{figure}
\plottwo{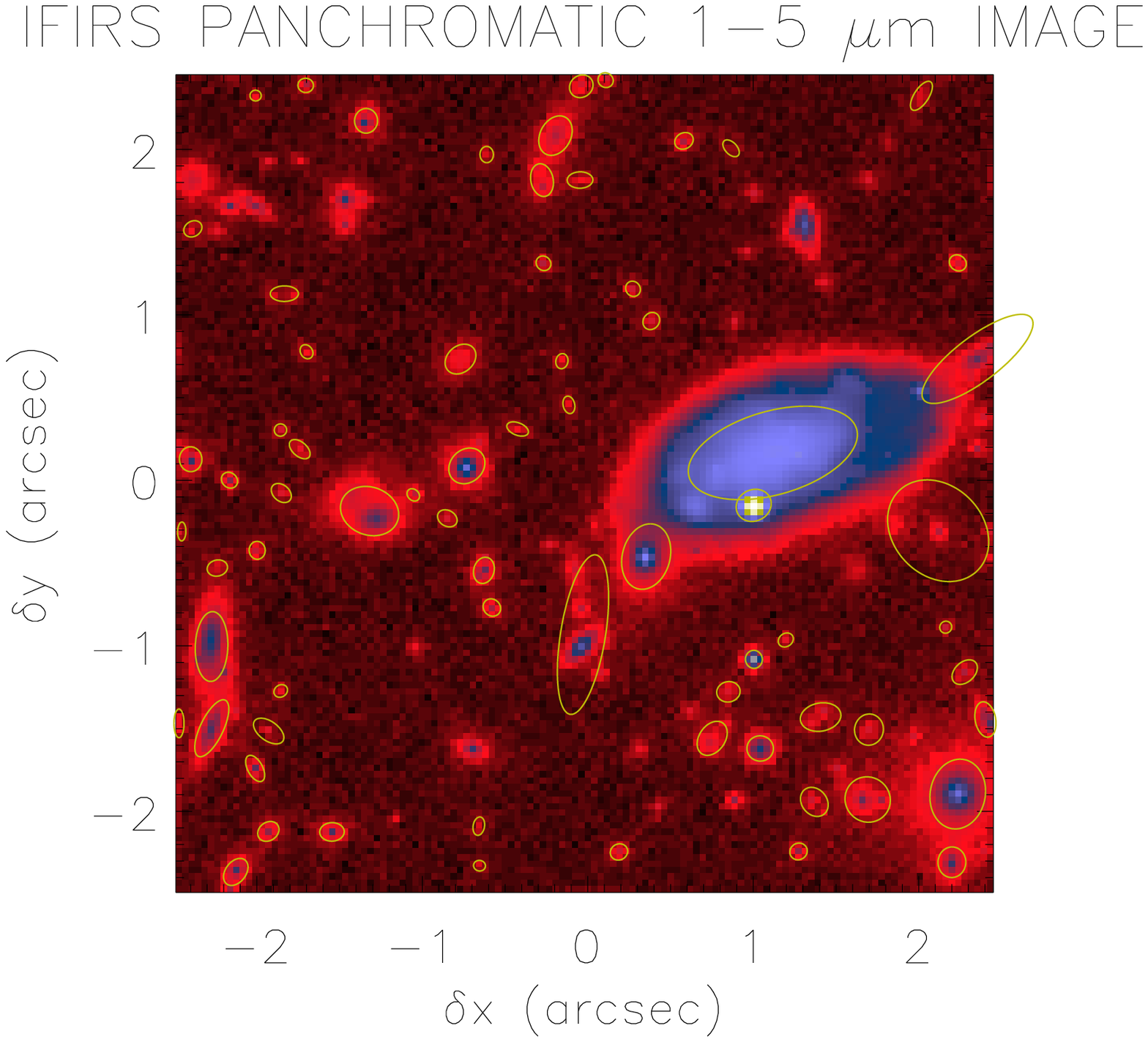}{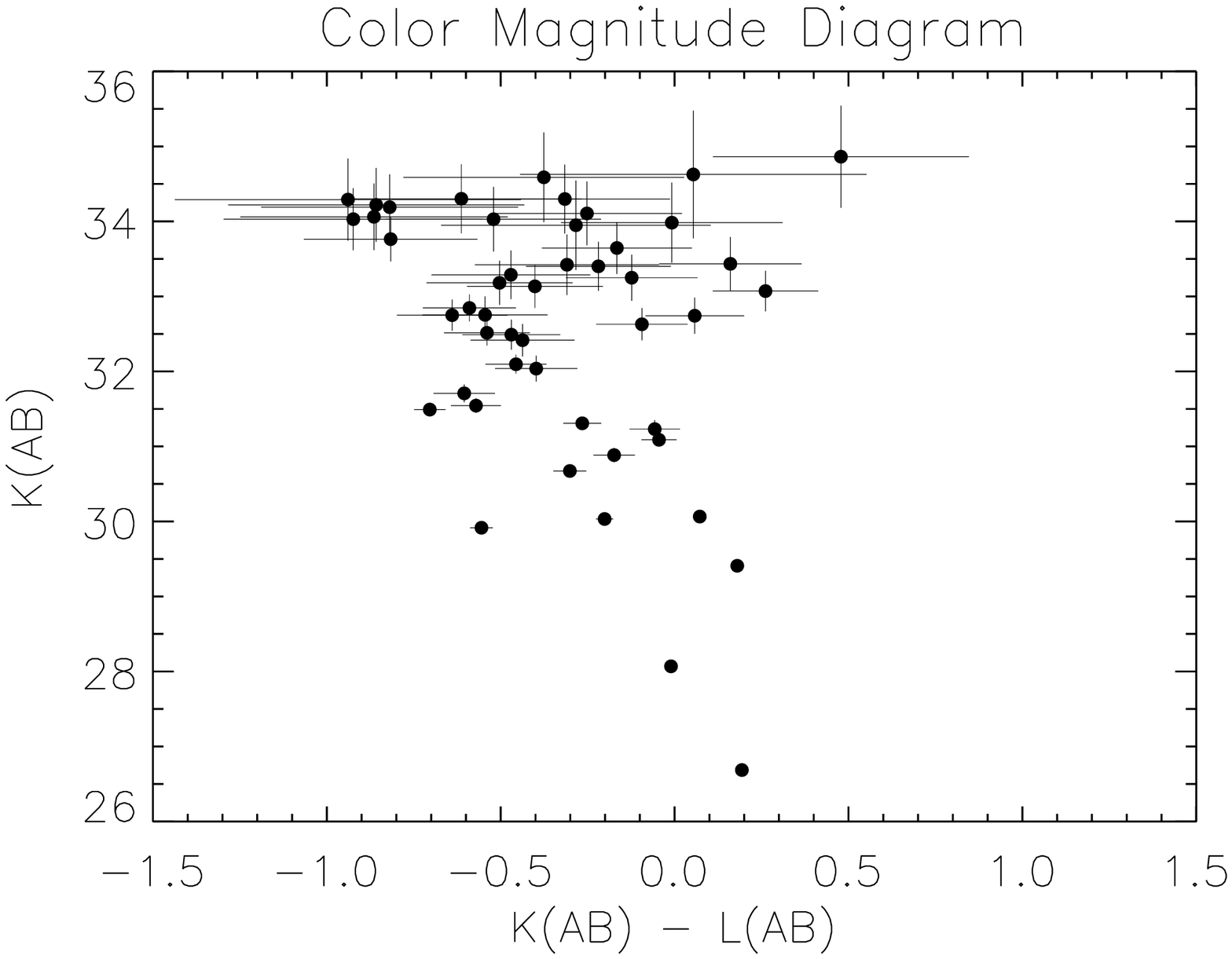}
\caption{Simulated ultra-deep ($10^6$~s) $ZJHKLM$
observations.  This region corresponds to 0.02\%
of the IFIRS field of view and contains a $K = 26.0$~AB, $z=4.67$
spiral and 60 $z>5$ star-forming galaxies brighter than $K\approx
33.5$~AB mag.  The left panel shows the panchromatic image, which has
been used by SExtractor program (Bertin et al. 1996) as the source
identification template. The corresponding $K-L/K$ color
magnitude diagram is shown on the right.}
\label{cmd}
\end{figure}

Fig.~\ref{cmd} shows the simulated data typical of ultra-deep low
resolution spectrophotometry of a $4.''9\times4.''9$ region
(0.02\% of the IFIRS FOV) with input data from Im \& Stockman (1998).
Crowding is significant: at least 6 of the $z>5$ galaxies lie behind
the spiral, and several others are obscured by other foreground
galaxies.  We have added a SN~Ia at the redshift of the spiral and a
$z=12$ star forming proto-cluster.  The proto-cluster has $F_\nu = 2$
nJy, and has been forming stars at a rate of 2.5 $M_\odot$ yr$^{-1}$
for $10^7$ yr.  In this example six broad photometric bands, $ZJHKLM$,
are synthesized.  The total exposure time of $10^6$~s is chosen to
match the ultra-deep exposure from the Galaxy Evolution DRM.  The 10
$\sigma$ limits are 32.7~AB at $K$ and 34.5~AB (panchromatic).

\begin{figure}[t]
\plotfiddle{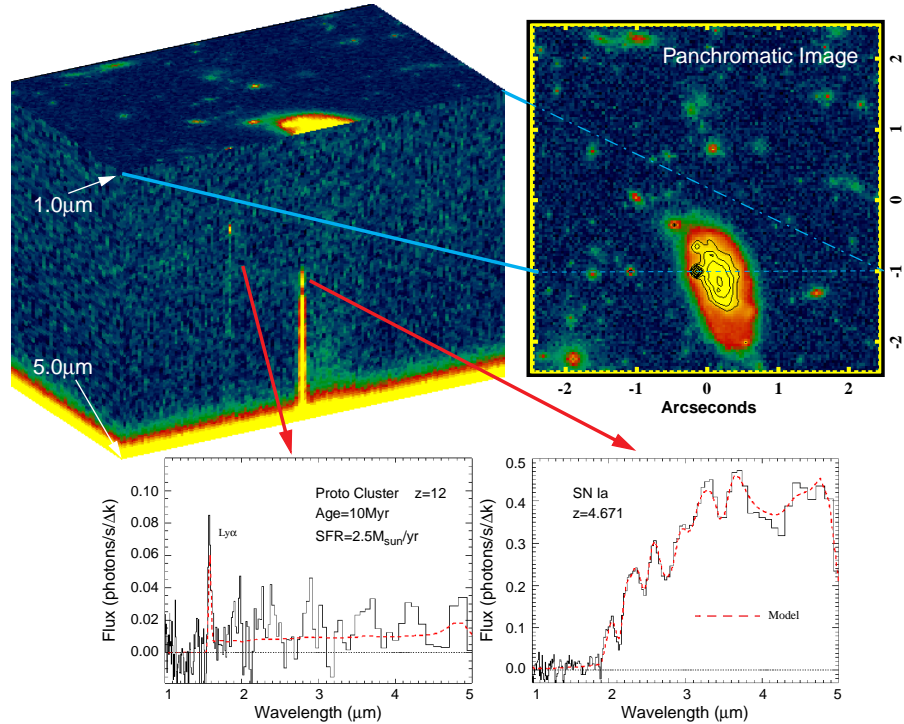}{3.5in}{-90}{45}{45}{-180}{270}
\caption{\small Simulated $10^5$~s $R=100$ $1-5\ \mu$m
IFIRS observation. The data cube, panchromatic image, and
spectral extractions are displayed.  A $4.''9\times4.''9$ (i.e.,
0.02\% of the IFIRS FOV) field is shown.  The input data is from Im \&
Stockman (1998).  The data cube (top left) has been sliced open to
reveal the spectra of a $z$=12 star forming proto-cluster and a
SN~Ia $z$=4.67 (the redshift of the spiral galaxy).  
The horizontal dashed line
on the panchromatic image (top right),  shows the
where the cube is sliced. Extracted spectra 
are shown at the bottom 
along with the input model (dashed). 
The NEFD at 3~$\mu$m is $\simeq 3$~ nJy.}
\label{deep-field} 
\end{figure}

Fig.~\ref{deep-field} shows simulated data from a $10^5$~s, $R=100$,
$1-5$~$\mu$m observation.  The extracted IFIRS spectrum of the $z=12$
proto-cluster shows Ly$\alpha$ emission, the continuum longward of the
line, and the continuum break across the line due to intervening HI.
The detection of emission lines allows IFIRS to measure redshifts and
star formation rates.  Given the large number of objects that will be
simultaneously observed and the deep imaging and spectroscopy that
will be available for each object, it will be possible to map out the
co-evolution of the star formation and morphological evolutionary
history of galaxies.  All objects with $F_\nu > 0.15$~nJy are detected
in the panchromatic image.  The statistics of the deep image
illustrate the approach to the crowding limit and the diagnostic power
of panchromatic imaging: approximately 33\% of the pixels in the
panchromatic image have values $>5\sigma$ above sky.  The simulation
illustrates the speed with which IFIRS can implement imaging and
spectroscopic surveys, the high quality data that IFIRS will produce,
and the need for IFIRS's full field spectroscopic ability to
spectrally and spatially separate background objects from foreground
ones.

\subsubsection{Growth of Structure \& Clustering:}
\begin{figure}[t]
\plotfiddle{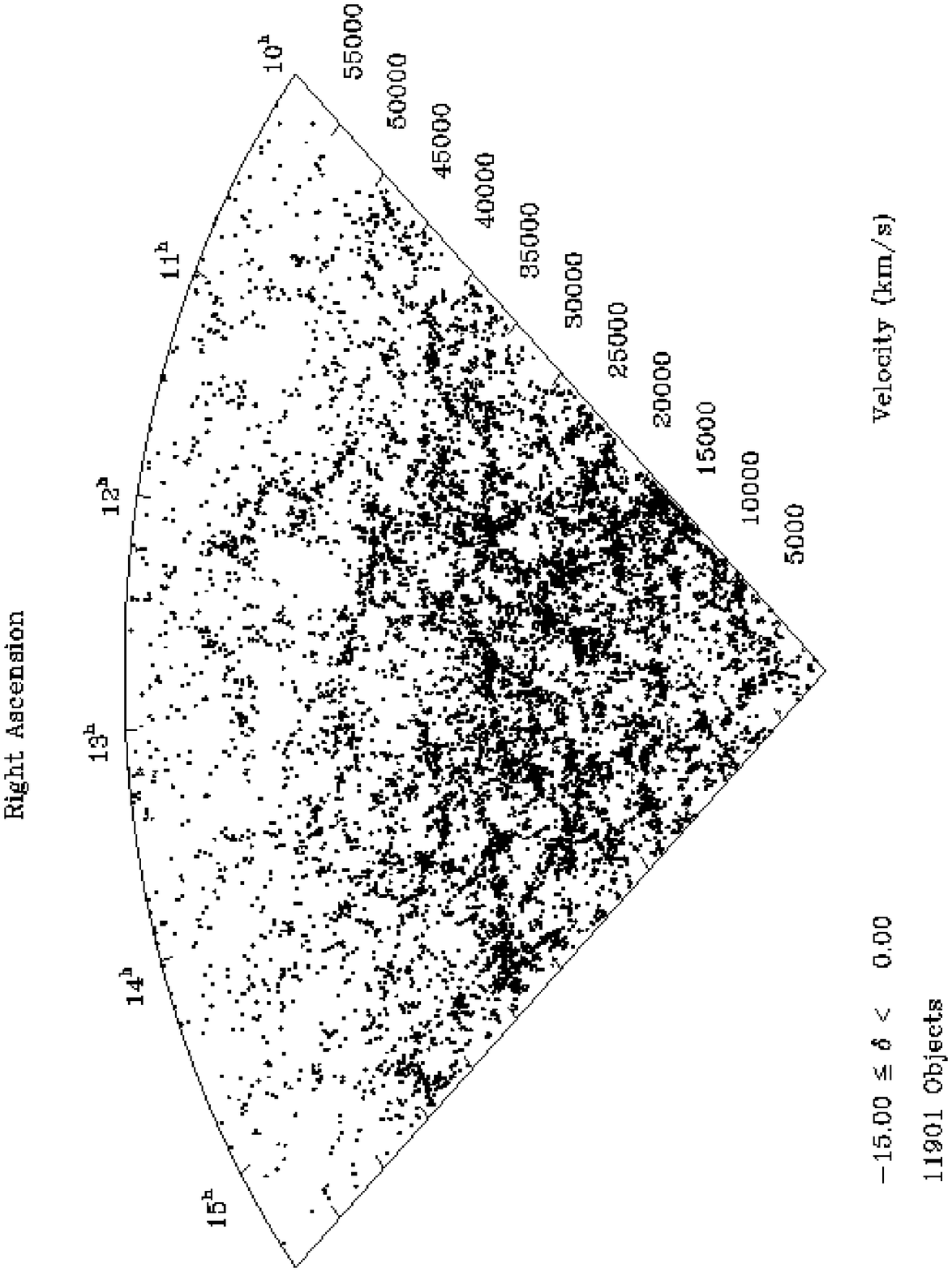}{1.0in}{-90}{20}{20}{-160}{90}
\plotfiddle{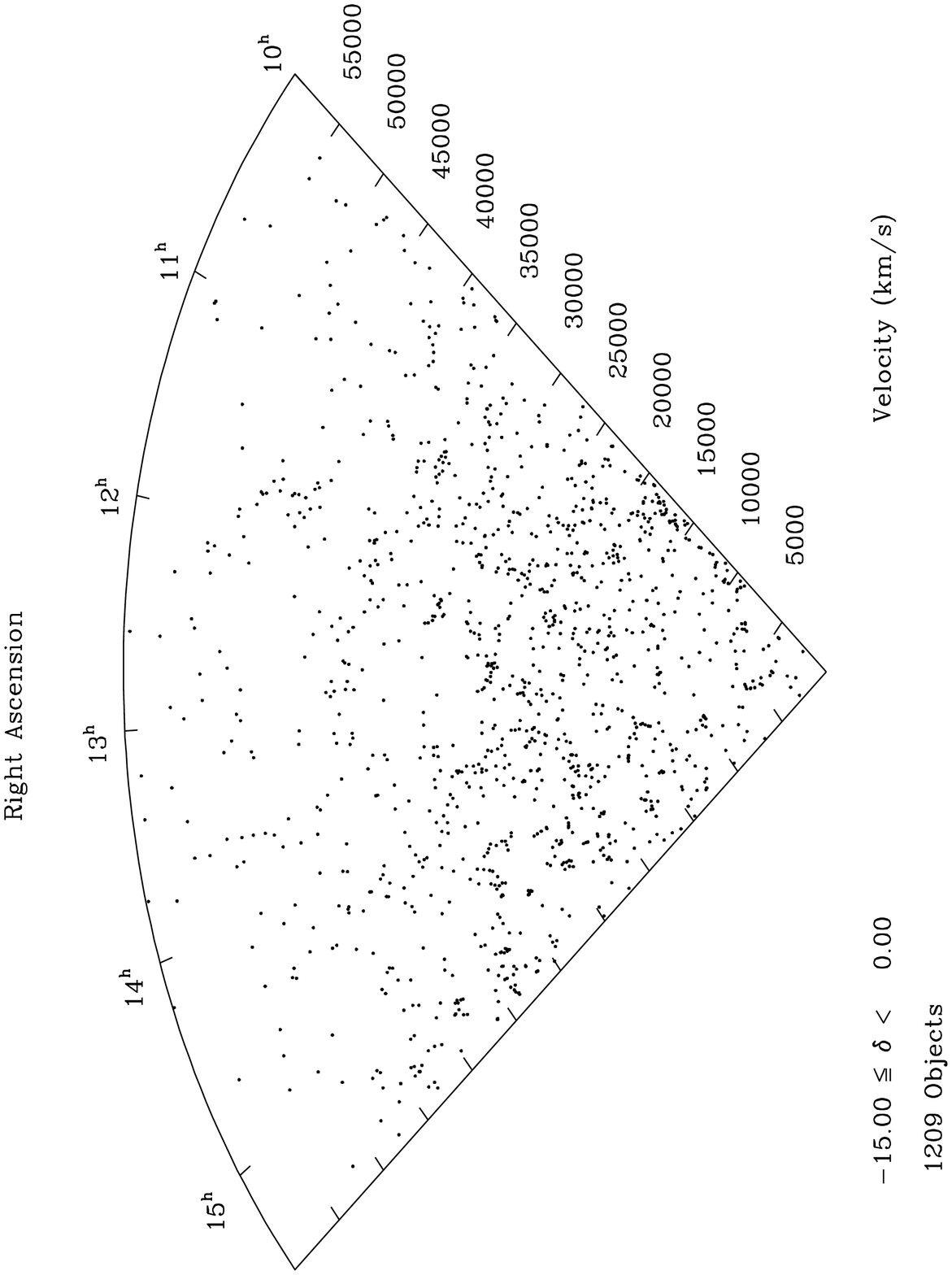}{0.0in}{-90}{20}{20}{-0}{115}
\caption{\small Two versions of a 15$^\circ$ slice from the Las
Campanas Redshift Survey (Schectman et al. 1996).  
In the left had panel all 11901 galaxies are plotted,
and structure is apparent. In the right hand panel only one galaxy in
ten (1209), selected at random, are plotted, and the structure is no
longer apparent (M. Postman, private communication).  IFIRS will
produce redshifts for tens of thousands of galaxies per field over a vastly
greater redshift range, rendering NGST extremely effective for
studying the evolution of structure.}
\label{clustering}
\end{figure}

The redshift evolution of galaxy clustering is a fundamental test of
structure formation theories. The amount and form of clustering in a
galaxy population can be estimated from a redshift survey (e.g.,
Fig.~\ref{clustering}) using a variety of sophisticated statistical
measures, the simplest of which is the two-point correlation function
$\xi$. Formal error estimates for $\xi$ scale as the square root of
the number of galaxies in each redshift bin, so very large samples
are needed.  Samples of a few hundred galaxies resulting from
conventional MOS surveys in a particular field will be entirely
inadequate to address this problem. The need for large samples is made
more extreme by the vastly extended redshift range probed by NGST and
by the clear desire to subdivide any galaxy sample still further to
look for clustering as a function of galaxy properties such as
luminosity, color, star formation rate, or morphology (cf. Kauffmann
et al. 1999).  For such studies, sample sizes of $\sim 10^5$ objects,
much larger than estimated in the Galaxy DRM proposals, are needed and
feasible with the spatial multiplexing capability of IFIRS.

Since IFIRS deep field observations produce large numbers of redshifts
in a single field ($1.8 \times 1.8$~Mpc at $z=5$), 
isolating correlated structures (clusters,
filaments, sheets) in redshift space is automatic.  Hence, IFIRS will
enable the investigation of various relations (e.g., morphology
vs. density, star formation rate vs. density) in coherent structures
and trace the evolution of these relations as a function of
redshift. IFIRS surveys in the regions of rich galaxy clusters will
provide ready identification of cluster members and diffuse line
emission associated with the clusters, as well as the ability to
identify and obtain redshifts for background lensed objects.

\begin{figure}[t]
\plotfiddle{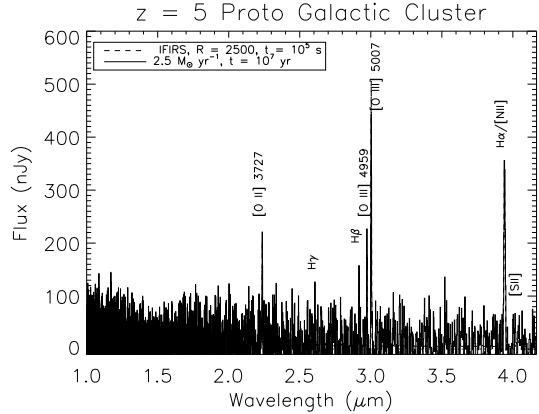}{2.5in}{0}{35}{35}{-160}{0}
\caption{\small Simulated $R = 2500$ (pure FTS mode) observation. The
target is a $K = 29.2$~AB star forming galaxy ($0.25 L^*$) at $z=5$.
[O II], [O III], H$\beta$ and H$\alpha$/[NII] are detected. The H$\beta$,
[O III] $\lambda\lambda$ 4959, 5007 complex is resolved permitting
measurement of the enrichment of $\alpha$-process elements.  The
rest frame equivalent width of [OII] is 40 \AA.  }
\label{z5gal2000}
\end{figure}

Fig.~\ref{z5gal2000} shows that pure FTS observations are appropriate
for high spectral resolution observations needed to study clustering
properties. Since the sensitivity of an IFTS to an unresolved emission
line is independent of spectral resolution IFIRS has sufficient
sensitivity to detect emission lines ($EW \simeq 20$~\AA) in faint
($K\simeq 29$~AB) galaxies. This permits IFIRS to study kinematics,
chemical composition, or peculiar velocities.

\section{CONCLUSIONS}

IFIRS enables a wide variety of NGST science, while it lends to
spectroscopy the potential for serendipity that is normally accorded
only to imaging.  Thus, it is a powerful tool for discovery.  The
advantages of the IFTS concept are:

\begin{enumerate}
\itemsep=-5pt
\item Deep imaging acquired simultaneously with higher spectral
resolution data over a broad wavelength range.
\item ``Hands-off'', unbiased, multi-object, slitless spectroscopy
(ideal for moving objects).  Efficient in confusion limit.
\item Flexible resolution ($R=1-10,000$).
\item High throughput (near 100\%) dual-port design.
\item Tolerant of cosmic rays, read-noise, dark current, and light leaks.
\item Simple and reliable calibration.  
\item Proven technology with low development costs.
\end{enumerate}

\noindent
I am indebted to the IFIRS team for their contributions: M. Abrams,
C. Bennett, J. Carr, K. Cook, A. Dey, R. Hertel, N. Macoy, S. Morris,
J. Najita, A. Villemaire, E. Wishnow, \& R. Wurtz.  This work was
support by NASA.


\begin{references}

\reference Beer, R. 1992, Remote Sensing by Fourier Transform
Spectrometry , John Wiley and Sons, New York, N. Y.

\reference Bennett, C. L. 2000, in Imaging the Universe in Three
Dimensions. ASP Conference Series, W. van Breugel \& J. Bland-Hawthorn
(eds.)

\reference Bertin, E. \& 
Arnouts, S. 1996, \aaps, 117, 393 

\reference Graham, J.R. et al. 1998, PASP, 110, 1205

\reference Graham, J. R. 2000, in Imaging the Universe in Three
Dimensions. ASP Conference Series, W. van Breugel \& J. Bland-Hawthorn
(eds.)

\reference Haiman, Z.  \& Loeb, A. 1998, ApJ, 503, 505 

\reference Im, M. and Stockman, H.S. 1998, in Science with the NGST , eds.,
E.P. Smith and A. Koratkar, ASP Conf. Ser. vol 133, p. 263

\reference Kauffmann, G., Colberg, J.M., Diaferio, A., and White, S.D.M. 1999,
MNRAS, 303, 188

\reference Shectman, S. A. et al. 1996, \apj, 470, 172 

\end{references}
\end{document}